\documentclass[aps,pre,preprint,showpacs,endfloats*]{revtex4-1}

\usepackage{amssymb,amsmath}
\usepackage{graphicx}
\usepackage{color}
\usepackage{bm}

\newcommand{\rmn}[1]{\mathrm{#1}}
\newcommand{\mean}[1] {\overline{ #1}}
\newcommand{\meanlong}[1] {\langle #1\rangle}
\newcommand{\prm}{^{\prime}}
\newcommand{\conv}{\!*}

\begin{document}

\bibliographystyle{apsrev}
\title{Growth rate of Rayleigh-Taylor turbulent mixing layers \\ with the foliation approach}
\author{\surname{Olivier} Poujade$^{*}$}
\author{\surname{Mathieu} Peybernes}

\affiliation{CEA, DAM, DIF, F-91297 Arpajon, France }

\date{\today}


\begin{abstract}
For years, astrophysicists, plasma fusion and fluid physicists have puzzled over Rayleigh-Taylor turbulent mixing 
layers. In particular, strong discrepancies in the growth rates have been observed between experiments and numerical simulations. Although two phenomenological
mechanisms (mode-coupling and mode-competition) have brought some insight on these differences, convincing theoretical arguments are
missing to explain the observed values. In this paper, we provide an analytical expression of the growth rate compatible with both
mechanisms and is valide for a self-similar, low Atwood Rayleigh-Taylor turbulent mixing subjected to a constant or time-varying acceleration. The key
step in this work is the introduction of {\it foliated} averages and {\it foliated} turbulent spectra highlighted in our three dimensional numerical
simulations. We show that the exact value of the Rayleigh-Taylor growth rate not only depends upon the acceleration history but is also bound to the
power-law exponent of the {\it foliated} spectra at large scales. 
\end{abstract}

 \maketitle

\section{Introduction}

If allowed to evolve long enough, the Reynolds number of a Rayleigh-Taylor (RT) flow
\cite{rayl, tayl} increases and the mixing eventually becomes fully turbulent
\cite{alphag, cookcab}. Then, at late
time, the width of the mixing layer grows according to 
\begin{equation}
L(t)=\alpha_p\,{\cal A}\,g(t)\,t^2\label{width} 
\end{equation} when the gravitational field history \cite{al} is of the form
$g(t)\propto t^p$. The Atwood number  ${\cal A}$ quantifies the
density contrast of the initially pure fluids and is tied to the acceleration in any
buoyant flow. From a purely theoretical point of view, the  RT growth
rate coefficient $\alpha_p$ should be universal if the late stage of
the evolution of the mixing zone is self similar and should only
depend upon the exponent $p$ of the gravitational field
history. However, even today, the value of "alpha" ($\alpha$), which
is the name given by the "alpha-group collaboration" \cite{alphag} to
$\alpha_p/2$ when $p=0$ (constant acceleration),  is still a subject of
controversy. 

\par

For years, the discrepancy between its value inferred
from numerical simulations \cite{alphag, cook3, cab1, cookcab} (initialised with small wave
lengths), scattered around $0.05$, and its experimental values
\cite{dim1, dim15, and1, banerandrews}, twice as large, has thrown doubt on its universality. Not
long ago, the heuristic mode-competition mechanism \cite{dim3, dim4}
put an end to this paradox: adding a small amount of modes of large
wave length to the perturbation of the initial interface
helps increase the value of  $\alpha$ and reconciles
numerical simulations with laboratory experiments. But, it is also
instrumental in demonstrating that the initial conditions in any
laboratory experiments do feature these modes of large wave length
triggered by external causes like vibrations and residual motions. Flows of interest
(geophysics, atmospheric, astrophysics \cite{zing, sc1}), to a large extent, are not
confined to a laboratory. Orders of magnitude may separate the length
scales of the interface perturbation with the geometric size of the
flow. This is why, the mode-coupling mechanism, initiated with small wave length modes, is thought to be
realised in astrophysical mixing \cite{cookcab} and other natural flows. 

\par

Here, in order to find the theoretical value of the growth rate ($\alpha_p$), we have adopted a point of view which has
been used in the study of homogeneous turbulence (HT) for almost
half a century but never investigated in RT turbulent mixing. The
"large-scale structure of homogeneous turbulence" \cite{batch,saff}
has been studied to understand the conjectured persistence of big eddies
and the decay rate of various physical quantities in turbulent
flows. At large scales, the turbulent kinetic energy density spectrum
behaves as $E(k)\propto k^s$ where the value of $s$ depends on the
initialisation of freely decaying HT in a way which is still unclear
\cite{dav1, dav2, dav3, les1, les2}. Bulk quantities of such flows, like turbulent kinetic
energy, vary as power-laws at late time and their exponents appear solely related to the value of $s$ \cite{chas}. That underlines the fact that
large scale spectra have a strong influence on the behaviour of the
overall flow. The theoretical study of freely decaying HT is made difficult by the fact that large
scales are driven by non linear effect (triadic interaction). In contrast, the RT turbulence, although
anisotropic, displays a significant difference with freely decaying HT. When plunged in a gravitational
field, density fluctuations produce motion through buoyancy. This production mechanism affects all
scales \cite{op}, even the largest, and produces turbulent kinetic energy at all times. It may not be the dominant effect in the inertial range, as argued in \cite{chert} and \cite{boff}, but it overcomes
non linear transfer mechanism at large scales making the theoretical study of RT turbulence much easier. 

\par

Here we demonstrate that it is possible to display an exact formula for $\alpha_p$ which
depends upon $p$, the exponent of the acceleration history, and $s$, the power-law exponent at large scales
of {\it foliated} spectra that rise up naturally, along with {\it foliated} average, from the theoretical developments hereafter. This {\it foliation} procedure is introduced for the first time in turbulent mixing theory because, among other interesting features, it allows to cancel out the formal effect of pressure in the equations, while keeping its physical effect on the whole mixing zone, making theoretical calculations amenable.  

\section{Transverse average}

A RT mixing occurs whenever two fluids of different density, initially separated by a sharp interface, are subjected to a gravitational field directed from the heavier (density $\rho_h$) to the lighter (density $\rho_l$) and when the initial interface is slightly distorted by a perturbation whose spectrum contains modes with sufficiently large wave lengths to overcome viscous effects at the onset. Such a flow, with two incompressible fluids in the low Atwood limit ${\cal
  A}=(\rho_h-\rho_l)/(\rho_h+\rho_l)\ll 1$ (Boussinesq approximation),
is governed \cite{clarkristo} by a concentration equation (\ref{eqc}), the Navier Stokes
equation supplemented with a buoyant source term (\ref{equ}), and the
incompressibility constraint (\ref{eqdiv})
\begin{eqnarray}
\partial_t\,c+\left(\bm{u}\bm{\nabla}\right)c\,&=&\kappa\,\Delta c,\label{eqc}\\
\partial_t\,\bm{u}+\left(\bm{u}\bm{\nabla}\right)\bm{u}\,&=&-\bm{\nabla} P+2{\cal A}\,\bm{g}\,c+\nu\,\Delta\bm{u},\label{equ}\\
\bm{\nabla}\cdot\bm{u} &=&0,\label{eqdiv}
\end{eqnarray} where $c\left(\bm{x},t\right)$ is the
  mass fraction of the heavy fluid ($c=0$ in pure light fluid and $c=1$ in pure heavy fluid), ${\bm u}\left(\bm{x},t\right)$ 
  the velocity field and $p\left(\bm{x},t\right)$ the pressure field. The molecular diffusivity $\kappa$ and the kinematic viscosity of the mixture $\nu$ are both assumed to be constant. The gravitational acceleration vector
  $\bm{g}(t)$ points down and its intensity may be time dependent as outlined earlier. In our simulations (described in the section \S\ref{sec::sim}), the flow fills
  a rectangular box of size $H\times H\times 2H$ (see Fig.\ref{fig:fig1} for more details).

The calculation starts by averaging (\ref{eqc}-\ref{eqdiv}). The transverse spatial average of a physical quantity $q$ at height $z$ is
defined according to
\begin{equation}
\mean{q}\left(z,t\right)=\frac{1}{H^2}\int_{H^2}\:\rmn{d} x\,\rmn{d} y\,q\left(x,y,z,t\right)~,\label{meanperp}
\end{equation} along with its fluctuating part
$q\prm\left(x,y,z,t\right)=q\left(x,y,z,t\right)-\mean{q}\left(z,t\right)$. Velocity, concentration and pressure fields can be decomposed into mean and fluctuating parts: $\bm{u}=\mean{{\bm u}}+{\bm u}\prm$, $c=\mean{c}+c\prm$ and $p=\mean{p}+p\prm$. The velocity along the $z$ axis will be called $u_z$ and $\bm{v}$ will denote the transverse component in the $xy$ plane so that $\bm{u}=(\bm{v},u_z)$.
 The divergence
 equation (\ref{eqdiv}) and the fact that vertical velocity cancels at the top and bottom walls yield $\mean{u_z}\left(z,t\right)=0 $. Similarly, 
 in a transverse plane at height $z$, there is an equal probability of finding $v_i$ and $-v_i$ 
 (right/left symmetry in the experimental setup); thus $\mean{v_i}=0$, 
 $\mean{v_i\prm\, c\prm}=0$ and $\mean{v_i\prm\, u_z\prm}=0$ at all times
 \cite{clarkristo} ($i=x,\,y$). The behaviour of
the fluctuating flow will be studied at large scales (low $k$) and at high
Reynolds number. Viscous and diffusive terms (in
factor of $\nu$ and $\kappa$) can then be discarded because in the spectral domain, they contribute a factor $\nu\,k^2$ 
and $\kappa\, k^2$. These terms become important only after a time $t\sim 1/(\nu\, k^2)$. That happens after a time 
$H^2/\nu$, or $H^2/\kappa$, for viscous, or diffusive, terms at large scale $k\sim1/H$. This is well above 
$\sqrt{H/({\cal A}\,g)}$ (the time required by the RT mixing zone to grow and to fill the domain of width $H$) 
since $H$ is well above $\Delta\sim (\nu^2/{\cal A}\,g)^{1/3}$ ($\Delta$ being the mesh size as in \cite{alphag}). 
Therefore, the resulting set of equations governing the evolution of the fluctuating flow at large scales reads
\begin{eqnarray}
&&\partial_t c\prm+u_z\prm\partial_z\mean{c}+\partial_i\left(v_i\prm\,c\prm\right)+\partial_z\left(u_z\prm\,c\prm\right)=\nonumber\\
&&\partial_z\mean{u_z\prm\,c\prm},\label{eqfc}\\
&&\partial_t u_z\prm+\partial_j\left(v_j\prm\,u_z\prm\right)+\partial_z\left(u_z\prm\,u_z\prm\right)-\partial_z\left(\mean{u_z\prm\,u_z\prm}\right)=\nonumber\\
&&-\partial_z p\prm-2{\cal
 A}\,g\,c\prm,\label{eqfuz}\\
&&\partial_t
v_i\prm+\partial_j\left(v_j\prm\,v_i\prm\right)+\partial_z\left(u_z\prm\,v_i\prm\right)-\partial_j\left(\mean{v_j\prm\,v_i\prm}\right)=\nonumber\\
&&-\partial_i\,p\prm,\label{eqfux}
\end{eqnarray} where Einstein summation convention is used on the repeated
indices $i$ and $j\in\{x,y\}$.

\section{Foliated average}

The last equations can be Fourier transformed for the purpose of showing that the growth rate of a RT mixing zone depends upon the structure of turbulence
at large scales. Since the physical domain is assumed to be periodic along $x$ and $y$ (see Fig.\ref{fig:fig1}), one defines a Fourier transform operator
acting on these two directions only (transverse directions). It turns a generic quantity $f\left(\bm{r},z,t\right)$,  where $\bm{r}=(x,y)$ is the
transverse position, into $\widetilde{f}\left(\bm{k},z,t\right)$, where $\bm{k}=(k_x,k_y)$ is the transverse wave vector. Equations (\ref{eqfc}-\ref{eqfux})  become
\begin{eqnarray}
\partial_t \widetilde{c\prm}&=&-\widetilde{u_z\prm}\,\partial_z\mean{c}-\partial_z(\widetilde{u\prm_z}\conv\widetilde{c\prm})\nonumber\\
&&-\imath\,k_j\,\widetilde{v\prm_j}\conv\widetilde{c\prm},\label{eqffc}\\
\partial_t\widetilde{u_z\prm}&=&-\partial_z(\widetilde{u\prm_z}\conv\widetilde{u\prm_z})-\partial_z
\widetilde{p\prm}-2{\cal A}\,g\,\widetilde{c\prm}\nonumber\\
&&-\imath\,k_j\,\widetilde{v\prm_j}\conv\widetilde{u\prm_z},\label{eqffuz}\\
\partial_t\widetilde{v\prm_i}&=&-\partial_z(\widetilde{u\prm_z}\conv\widetilde{v\prm_i})-\imath\,
	{k_i}\,\widetilde{p\prm}\nonumber\\
&&-\imath\,k_j\,\widetilde{v\prm_j}\conv\widetilde{v\prm_i},\label{eqffv}
\end{eqnarray} where $*$ is the usual folded product.
 In order to simplify (\ref{eqffc}-\ref{eqffv}) one can eliminate the effect of
 $\partial_z$, and with it, the effect of pressure and non locality, by
 integrating these equations along the anisotropic direction $z$. Therefore,
 these arguments suggest introducing the {\it foliated} spatial average
 which, for a physical quantity $q$, is defined according to
\begin{eqnarray}
\meanlong{q}\left(x,y,t\right)&=&\frac{1}{H}\,\int\,q\left(x,y,z,t\right)\:\rmn{d} z~.\label{meanlong}
\end{eqnarray} It amounts to chopping the domain (foliation) in elementary slices,
transverse to the anisotropic direction, and to squeezing them by adding them
up, resulting in an effective 2D flow (see Fig.\ref{fig:fig2}). Both in the concentration
equation (\ref{eqffc}) and in the equation of motion (\ref{eqffuz}),
$\meanlong{\partial_z(\widetilde{u\prm_z}\conv\widetilde{q\prm})}=0$ (where $q\prm=c\prm$ or
$u_z\prm$) because at top and bottom walls $\widetilde{u\prm_z}=0$. The average of the pressure
gradient is $\meanlong{\partial_z \widetilde{p\prm}}= 0$ because pressure
fluctuations decrease outside the mixing zone. It has to be assumed that the
top and bottom walls are far enough so that
$\lim_{z\to\pm\infty}\widetilde{p\prm}(z,\bm{k},t)=0$. Consequently, the
boundaries do not affect the flow within the mixing zone if $L(t)\ll 2\,H$
(which can be relaxed to $L(t)\lessapprox H$ in simulations). This is why, the resulting
concentration equation and equation of motion after {\it foliated} average are
\begin{eqnarray}
\partial_t \meanlong{\widetilde{c\prm}}&=&-\meanlong{\widetilde{u_z\prm}\,\partial_z\mean{c}\,}-\imath\,k_j\,\meanlong{\widetilde{v\prm_j}\conv\widetilde{c\prm}}~,\label{eocl}\\
\partial_t\meanlong{\widetilde{u_z\prm}}&=&-2{\cal  A}\,g\,\meanlong{\widetilde{c\prm}}-\imath\,k_j\,\meanlong{\widetilde{v\prm_j}\conv\widetilde{u\prm_z}}~.\label{eosl}
\end{eqnarray} In eq.(\ref{eocl}), the term $\meanlong{\widetilde{u_z\prm}\,\partial_z\mean{c}\,}$ can be simplified because, at small Atwood number, the profile of
$\mean{c}(z,t)$ deviates only from a straight line (constant slope) at the edges of the mixing zone. This approximation is corroborated experimentally
\cite{and1,banerandrews} and numerically, in our simulations (${\cal A}=0.1$), to within statistical fluctuations
due, at a given height $z$, to the finite number $N(t)\approx H^2/\ell(t)^2$ of eddies in the transverse plane of area $H^2$ (the integral length scale
$\ell(t)$, i.e. the diameter of a typical eddy, varies like $L(t)\propto
t^{p+2}$ in the self-similar regime and these statistical fluctuations become negligible in the limit $H\rightarrow +\infty$ since their rms amplitude is of the order
$1/\sqrt{N(t)}\approx \ell(t)/H$).  In this limit, the derivative $\partial_z\mean{c}$ is uniform and equal to $1/L(t)$ within the mixing
zone to a good approximation. Thus, the theoretical development suggests defining $L$ such that $\partial_z\mean{c}=1/L(t)$ which can be brought close to another definition of the
  width at low Atwood number (see the end of \S\ref{sec::sim}). Therefore, and this is the only approximation made in this development, it is possible to write
\begin{equation}  
\meanlong{\widetilde{u_z\prm}\,\partial_z\mean{c}\,}\approx 
\meanlong{\widetilde{u_z\prm}}\,\partial_z\mean{c}=\meanlong{\widetilde{u_z\prm}}/L(t). \label{approx}
\end{equation} The contribution of
the edges is negligible in the foliated average because it represents a small portion of the integration domain. As a consequence, (i) at large scale ($k\ll 2\pi/\ell(t)$), (ii) at low Atwood
(iii) at high Reynolds number and (iv) in the limit $H\rightarrow +\infty$ (i.e. $L(t)\ll H$), the evolution of {\it foliated} second moments can be derived. From eqs.(\ref{eocl}-\ref{eosl}), straightforward algebraic
manipulations provide 
\begin{eqnarray}
\partial_t(\meanlong{\widetilde{c\prm}}\meanlong{\widetilde{c\prm}}^*)&=&-\frac{2\,\mathrm{Re}[\meanlong{\widetilde{u_z\prm}}\meanlong{\widetilde{c\prm}}^*]}{L(t)}\nonumber\\
&&-k_j\,\xi_j^c~,\label{ee1}\\
\partial_t(\meanlong{\widetilde{u_z\prm}}\meanlong{\widetilde{u_z\prm}}^*)&=&-4\,{\cal
 A}\,g\,\mathrm{Re}[\meanlong{\widetilde{u_z\prm}}\meanlong{\widetilde{c\prm}}^*]\nonumber\\
&&-k_j\,\xi_j^z~,\label{ee2}\\
\partial_t(\mathrm{Re}[\meanlong{\widetilde{u_z\prm}}\meanlong{\widetilde{c\prm}}^*])&=&-2{\cal A}\,g\,\meanlong{\widetilde{c\prm}}\meanlong{\widetilde{c\prm}}^*-\frac{\meanlong{\widetilde{u_z\prm}}\meanlong{\widetilde{u_z\prm}}^*}{L(t)}\nonumber\\
&&-k_j\,\xi_j^{cz}~,\label{ee3}
\end{eqnarray} where $\xi_j^c=2\,\mathrm{Re}[\imath\meanlong{\widetilde{v_j\prm}*\widetilde{c\prm}}\meanlong{\widetilde{c\prm}}^*]$,
$\xi_j^z=2\,\mathrm{Re}[\imath\meanlong{\widetilde{v_j\prm}*\widetilde{u_z\prm}}\meanlong{\widetilde{u_z\prm}}^*]$ and
$\xi_j^{cz}=\mathrm{Re}[\imath\meanlong{\widetilde{v_j\prm}*\widetilde{u_z\prm}}\meanlong{\widetilde{c\prm}}^*]+\mathrm{Im}[\imath\meanlong{\widetilde{v_j\prm}*\widetilde{c\prm}}\meanlong{\widetilde{u_z\prm}}^*]$
cancelled out when ensemble averaged. 

\section{Ensemble averaged foliated spectra}

Let us note $\widehat{q}$ the ensemble average of a generic quantity $q$. Different realizations of the RT flow 
have the same initial spectrum of interface perturbations but the initial phases of the modes are different 
and are generated at random. As a result, if one realization is initialized with a set of random phases, called 
$\Phi(\bm{k},t=0)$, and produces $\xi_j(\bm{k},t)$, then the particular realization 
initialized with phases $\Phi(\bm{k},t=0)+\pi$ will produce $-\xi_j(\bm{k},t)$ with the exact same 
probability. When averaged, these two contributions cancel out and therefore the ensemble average over all 
possible realizations---or equivalently, over all possible initializations---translates into $\widehat{\xi_j^c}(\bm{k},t)=0$, 
$\widehat{\xi_j^z}(\bm{k},t)=0$ and $\widehat{\xi_j^{cz}}(\bm{k},t)=0$. 

It suggests defining the ensemble
averaged {\it foliated} spectra for kinetic energy, concentration and
production respectively as 
\begin{eqnarray}
{\cal E}_z(k,\,t)&=&\frac{k}{2}\,\int\,\mathrm{d}\Omega\,\widehat{\meanlong{\widetilde{u_z\prm}}\meanlong{\widetilde{u_z\prm}^*}}~,\\
{\cal E}_c(k,\,t)&=&k\,\int\,\mathrm{d}\Omega\,\widehat{\meanlong{\widetilde{c\prm}}\meanlong{\widetilde{c\prm}^*}}~,\\
{\cal E}_{cz}(k,\,t)&=&-k\,\int\,\mathrm{d}\Omega\,\mathrm{Re}[\widehat{\meanlong{\widetilde{u_z\prm}}\meanlong{\widetilde{c\prm}^*}]}~,
\end{eqnarray} where the integration over the solid angle $\mathrm{d}\Omega$ is to be understood as the
integration over all directions of $\bm{k}$ (in the transverse
plane).  The $\xi$s in eqs. (\ref{ee1}-\ref{ee3}) disapear when ensemble averaged and the ensemble averaged {\it foliated} spectra, referred to as {\it foliated} spectra for short, are coupled to each other at large scales
according to 
\begin{eqnarray}
\dot{{\cal E}}_c&=&2\,\frac{{\cal E}_{cz}}{L(t)}~,\label{fin1}\\
\dot{{\cal E}}_z&=&2\,{\cal A}\,g\,{\cal E}_{cz}~,\label{fin2}\\
\dot{{\cal E}}_{cz}&=&2{\cal A}\,g\,{\cal E}_c+2\frac{{\cal E}_z}{L(t)}\label{fin3}~.
\end{eqnarray} It should be stressed that this closed set of equations does not hold if {\it foliated} average
operators are removed or even replaced by transverse average because of the non local effect of pressure. Since these equations are linear, autonomous in $k$ and do not display derivatives with respect
to $k$, it comes as a byproduct prediction of the theory that, with the aforementioned definitions, all three {\it foliated} spectra must have the same power-law exponent at large scales:
$k^s$. The exponent $s$ is independent of time to comply with self-similarity. This common
value is a special feature of the {\it foliated} spectra (see Fig.\ref{fig:fig3}). {\it Foliated} spectra have
deliberately been designated by script ${\cal E}$ in order to stress the difference with
transverse spectra classically defined by 
\begin{eqnarray}
E_z(k,\,z,\,t)&=&\frac{k}{2}\,\int\,\mathrm{d}\Omega\,\widetilde{u_z\prm}\,\widetilde{u_z\prm}^*~,\\
E_c(k,\,z,\,t)&=&k\,\int\,\mathrm{d}\Omega\,\widetilde{c\prm}\,\widetilde{c\prm}^*~,\\
E_{cz}(k,\,z,\,t)&=&-k\,\int\,\mathrm{d}\Omega\,\mathrm{Re}[\widetilde{u_z\prm}\,\widetilde{c\prm}^*]~.
\end{eqnarray}  Both types of spectrum were compared to show that this common
exponent is specific to the {\it foliated} spectra whereas transverse
spectra (concentration and velocity for instance) have different power-law exponents \cite{cab1} at large scales. This can be
understood when realising that a transverse spectrum is calculated using Fourier modes on a slice at a
height $z$ in the middle of the mixing zone. A {\it foliated} spectrum, on the other
hand, uses the sum along $z$ of all these modes. The resulting spectrum is radically different
and benefits from interferences between modes at different heights.

\section{Growth rate formula}

For self-similarity to hold, the late time behaviour of the {\it foliated} spectra must be
\begin{eqnarray}
{\cal E}_z(k,t)&=&{\cal E}_z^{0}\,k^s\,t^{e_z}~,\label{fol1}\\
{\cal E}_c(k,t)&=&{\cal E}_c^{0}\,k^s\,t^{e_c}~,\\
{\cal E}_{cz}(k,t)&=&{\cal E}_{cz}^{0}\,k^s\,t^{e_{cz}}~\label{fol2},
\end{eqnarray} at low $k$ where ${\cal E}_c^{0}$, ${\cal E}_z^{0}$ and ${\cal E}_{cz}^{0}$ are constants independent from $k$ and $t$. In addition, the exponent of $k$ in each of  this averaged spectrum ($s$) and the time exponents  ($e_c,~e_z,~e_{cz}$)  must be constant in time. When replacing these expressions with ${\cal A}\,g(t)={\cal
  A}\,g_p\,t^p$ and $L(t)=\alpha_p\,{\cal A}\,g_p\,t^{p+2}$ in eqs. (\ref{fin1}-\ref{fin3}) one gets the following relations
\begin{eqnarray}
e_z&=&e_c+2(p+1)~,\label{resfol1}\\
e_{cz}&=&e_c+(p+1)~,\\
{\cal E}_z^{0}&=&(\alpha_p\,({\cal A}\,g_p)^2\,e_c/e_z)\,{\cal E}_c^{0}~,\\
{\cal E}_{cz}^{0}&=&(\alpha_p\,{\cal A}\,g_p\,e_c/2)\,{\cal E}_c^{0}~,\\
\alpha_p&=&\frac{8}{e_z\,e_c}~\label{resfol2},
\end{eqnarray} which come straight from Eqs.(\ref{eqc}-\ref{eqdiv}) within the approximation made in Eq.(\ref{approx}). Using
Eqs.(\ref{fol1}-\ref{fol2}) and eqs.(\ref{resfol1}-\ref{resfol2}), a non trivial exact result can be
deduced: ${\cal E}^2_{cz}(k,\,t)/\left(2\,{\cal
  E}_z(k,\,t){\cal  E}_c(k,\,t)\right)=1$ valid at large scale (see Fig.\ref{fig:fig4}). 

\par  
  
A fundamental feature of
concentration is that its fluctuation variance, $\mean{c^\prime c^\prime}$ tends to a constant in the
self-similar regime because concentration is a physical quantity bounded between  0 and 1. It cannot go to zero
for that would mean the mixing tends to be heterogeneous in the middle of the mixing zone, thereby going against the
fact that at all times any  side of the mixing zone must be supplied with pure fluid from the other
side. The mixed-fluid-supplying-channels have been highlighted in Fig.\ref{fig:fig1} and Fig.\ref{fig:fig2}. This is
exemplified by numerical simulations and experiments reporting molecular mixing rates ranging from
$0.7$ to $0.8$ \cite{alphag, cookcab, and1, clarkristo}. 

\par

The question now is: how about $\mean{\meanlong{c^\prime}\meanlong{c^\prime}}$? A very simple argument can be given: the discretized foliated average,
\begin{equation}
\meanlong{c^\prime}=\mathrm{d}z/H\,\sum_{i=-H/dz}^{H/dz}\,c^\prime_i~,
\end{equation} can be approximated by 
\begin{equation}
\meanlong{c^\prime}=\mathrm{d}z/H\,\sum_{i=-L(t)/dz}^{L(t)/dz}\,c^\prime_i~,
\end{equation} since $c^\prime$ is only non zero within the mixing zone. That is why,
\begin{eqnarray}
\meanlong{c^\prime}\meanlong{c^\prime}&\approx & \mathrm{d}z^2/H^2\,\sum_{i=-L(t)/dz}^{L(t)/dz}\sum_{j=-L(t)/dz}^{L(t)/dz}\,c^\prime_i c^\prime_j~,\\
\mean{\meanlong{c^\prime}\meanlong{c^\prime}}&\approx & \mathrm{d}z^2/H^2\,\sum_{i=-L(t)/dz}^{L(t)/dz}\sum_{j=-L(t)/dz}^{L(t)/dz}\,c_0^2\,\delta_{ij}~,
\end{eqnarray}  if we say $\mean{c^\prime c^\prime}=c_0^2$ in the self similar regime as described in the previous paragraph. Therefore,  
\begin{equation}\mean{\meanlong{c^\prime}\meanlong{c^\prime}}\approx \mathrm{d}z^2/H^2\,\sum_{i=-L(t)/dz}^{L(t)/dz}\,c_0^2=\mathrm{d}z/H^2\,L(t)\,c_0^2~.
\end{equation} As a result $\mean{\meanlong{c^\prime}\meanlong{c^\prime}}$ must grow
like $L(t)$ (corroborated by our numerical simulations). 

\par

Therefore, since $\mean{\meanlong{c^\prime}\meanlong{c^\prime}}=\int_0^{+\infty}\mathrm{d}k\,{\cal
E}_c(k,t)$ varies as $\int_0^{2\pi/\ell(t)\,\propto\, t^{-p-2}}\mathrm{d}k\,{\cal E}_c(k,t)\propto
t^{e_c-(p+2)(s+1)}$ and must evolve as $L(t)\propto t^{p+2}$, the value of $e_c$ is bound to be $(p+2)(s+2)$
and, by inference, the time exponents $e_z=(p+2)(s+2)+2(p+1)$ and
$e_{cz}=(p+2)(s+2)+(p+1)$. Therefore, the value of the
Rayleigh-Taylor growth rate for the turbulent mixing zone width,
$\alpha_p$, is shown to depend upon $p$, which is the acceleration exponent, and $s$,
the {\it foliated} spectrum power-law exponent at large scales, according to
\begin{equation}
\alpha_p(s)=\frac{8}{(p+2)(s+2)(p(s+4)+2(s+3))}~.\label{alphan}
\end{equation} This expression does not depend on $g_p$ as expected and confirmed in our simulations. 

\par

The demonstration leading to this formula did not provide any arguments against the idea of an
$s$ varying with $p$. However, growth rates, at various values of $p$ given by our simulations in
conjunction with those provided in \cite{youngsllor}, were fitted to the theoretical formula and a
remarkable collapse of the data was found for $s=4.0\pm 0.1$ assuming $s$ was independent of $p$
(see Fig.\ref{fig:fig5}). It is possible to check this result by using an other obvious method: direct inspection of the {\it foliated spectra}. The power-law at large scales of {\it foliated} spectra was checked on every simulation ($p=0$, $1$, $2$ and $3$) and was found to be compatible with $s=4$. The formula (\ref{alphan}) is therefore a predictive formula: it gives the value of $\alpha_p$ knowing $p$ (a controled parameter) and $s$ or, the other way around, it gives $s$ knowing $p$ and $\alpha_p$. Our theory did not provide a value for $s$ for it certainly depends upon the way the mixing flow reaches the self-similar regime. Moreover, RT turbulence in 2D is totally
different from RT in 3D \cite{chert, cel}. This dependance of $\alpha_p$ upon spatial dimension is hidden in the value of $s$. In order to find its value, the theory would have to cope with the transition regime to make the connexion with the linear growth of the initial conditions ($s=4$ seems to be compatible with annular spectrum initial conditions in 3D). It must be emphasized that the result (\ref{alphan}) is a proof that, in  the self-similar regime, the growth of the mixing zone only depends upon the structure of turbulence at large scales (appart from $p$, it depends on $s$ only).

\section{Implications}

Obviously, the formula for $\alpha_p$ carries information on the dynamics of the
mixing zone width. Not so obvious is the fact that it also provides information on the way this dynamic affects the mixing at
large scale. If it is assumed that the late time evolution of the mixing zone width does not depend
upon initial conditions - which may be true when the characteristic length scale of the initial perturbations
is much smaller than that of the physical domain $H$ - it is reasonable to admit that it must
depend on $L(t)$ (and its derivatives) and $g(t)$ which are the only control parameters left in
the problem. From this assumption, basis of the buoyancy-drag approach \cite{dim1,dim2}, it is possible
to build a simple evolution equation,
\begin{equation}
\ddot{L}(t)=C_b\,{\cal A}\,g(t)\,L(t)-C_d\,\dot{L}^2(t)/L(t)~,\label{bdf}
\end{equation} which depends on two adjustable constants. The constant $C_b$ quantifies buoyancy related to the
mixing since $C_b\,{\cal A}$ can be seen as an effective Atwood number: the fluids are not pure, with density
$\rho_l$ and $\rho_h$, in the mixing zone but they are partly mixed with intermediate densities and the smaller
$C_b$, the smaller the effective Atwood number, the stronger the mixing. The constant $C_d$ quantifies drag,
i.e. the exchange of momentum between raising and falling mixed-fluids structures. This phenomenological approach
does not by itself provide any relation between these two adjustable constants.
However, replacing the self-similar values (\ref{width}) of $L$ and $g$ in (\ref{bdf}) 
allows to recover the exact formula (\ref{alphan}) if and only if $C_b=4/(s+2)$ and $C_d=(s+2)/2$, which yields
$C_b=2/C_d$. This correspondence between the simple model and the exact result demonstrate the little importance
of small scale as opposed to large scale in the bulk dynamics of the mixing. Furthermore, it enables to
understand the influence of $s$ on "intuitive" physical mechanisms like buoyancy, through the effective Atwood
number, and drag. Accordingly, the bigger $s$, the stronger the drag,  the smaller the effective Atwood number
and the better the mixing. 

\par

This is in agreement with numerical results \cite{dim3,dim4} suggesting that a smaller proportion of
long wave length in the spectrum (bigger $s$: mode-coupling) decreases the growth rate of the mixing zone width
in opposition to a bigger proportion of long wave length (smaller $s$: mode-competition). Therefore, in the
framework of {\it foliated} spectra where $s$ is defined, the two phenomenological mechanisms of mode-coupling and mode-competition in Rayleigh-Taylor turbulent mixing can be explained and brought together. 

\par

There is no reason in
principle why techniques developed in this work could not be applied to other mixing flows with one anisotropic direction. For instance, in the case of
Richtmyer-Meshkov (RM) turbulence \cite{dim2}, ${\cal A}\,g=0$ and the mixing width $L(t)$ varies like $t^{\theta}$ in eqs.(\ref{fin1}-\ref{fin3}). Persistence of big eddies \cite{batch} is then an exact result  for the {\it foliated} RM flow since the rhs of eq.(\ref{fin2}) vanishes. The same type of reasoning that
led to eq.(\ref{alphan}) allows to find $\theta=2/(s+4)$ (also an exact result when $s$ is the power-law exponent of ensemble averaged {\it foliated}
spectra at large scale) which for $3\leq s\leq 4$ predicts a $1/4=0.250\leq\theta\leq 0.286=2/7$, in agreement with experimental and numerical results.

\section{Simulations}\label{sec::sim}

A set of four simulations of Rayleigh-Taylor turbulent mixing flows were carried out with the
incompressible code SURFER \cite{surfer} for different acceleration histories $g(t)\propto t^p$ with $p=0$,
$p=1$, $p=2$ and $p=3$. For our purpose, gravity was added to the original version. 

\par

An incompressible code is valuable to investigate flows under variable
acceleration because hydrostatic pressure balances instantaneously with the
acceleration field. On the contrary, a compressible simulation would have to
cope with the creation of acoustic waves that would travel at the
speed of sound to balance pressure with time-varying
acceleration. These acoustic waves would bounce on the
domain walls, would go back and forth through the simulation domain
and would be detrimental to the simulations. 

\par

Each of those simulations was performed with the same grid resolution: $128\times 128\times 256$ (along $x$, $y$ and $z$ respectively). The large scale outcomes of a numerical simulation of a Rayleigh-Taylor turbulent flow do not vary significantly, for a given type of initial condition, by increasing the resolution. This explains the choice of such a modest resolution allowing to carry out more simulations. 

\par

SURFER is parallelised, 3D and evolves two immiscible fluids separated by an
interface. To reconstruct and advance the
fluid interfaces in time, SURFER uses an exactly volume
conserving variant of the Volume of Fluid algorithm with a Piecewise Linear Interface Calculation method (VOF/PLIC) \cite{guey}. The density of
each fluid $\rho_h$ (for heavy) and $\rho_l$ (for light fluid) are
constant in time. The Navier-Stokes equation governing the
evolution of the velocity field $\bm{u}$ is given by :
\begin{equation}\label{NS}
\rho\,\frac{d\bm{u}}{dt} =-\bm{\nabla}p\,+ \bm{\nabla}\cdot (\eta\,\bm{S})\,+\,\sigma\,r\,\delta_S\,\bm{n} + \rho\,\bm{g}~,
\end{equation} where $\rho$ is the density, $p$ is the pressure and $\bm{g}$ is the acceleration. The dynamic viscosity $\eta$ equals $\eta_h=\rho_h\,\nu$ or
$\eta_l=\rho_l\,\nu$ (where $\nu$ is a common kinematic viscosity) and $\bm{S}$ is the rate of strain tensor defined by $S_{ij}=\partial_i\,u_j+\partial_j\,u_i$. The surface tension $\sigma$ depends on the particular two fluids that will be
heterogeneously mixed and the delta function, $\delta_S$, is
concentrated on the surface of the interface, $r$ is the mean curvature of this surface
and $\bm{n}$ is the unit normal on the surface. The discretisation of (\ref{NS}) is performed on a MAC-type
staggered grid and the pressure is computed using an iterative multigrid Poisson
solver.

\par

\begin{figure}[t]
\centering
\includegraphics[width=10cm]{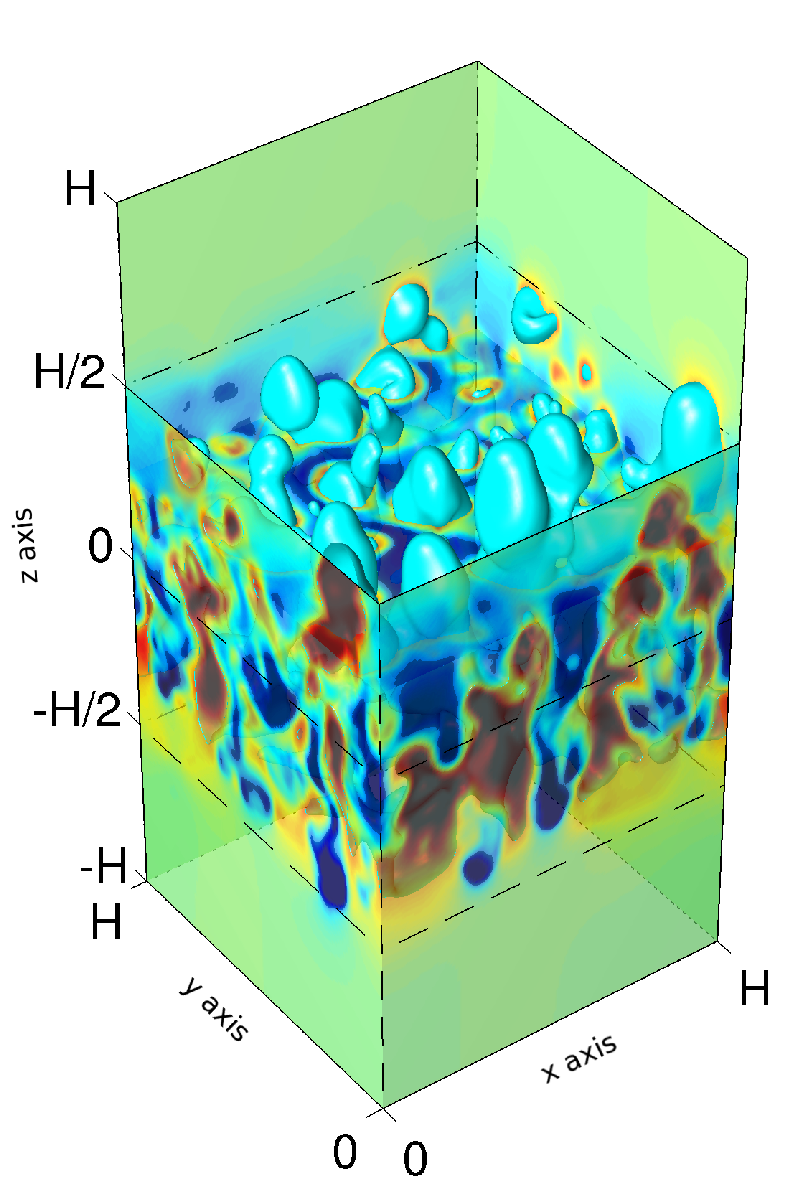}
\caption{(Color online) Image of a turbulent Rayleigh-Taylor mixing zone for a constant acceleration ($p=0$) when $L(t)=H$
 from our numerical simulations. The axes are defined in such a way that the mixing zone grows along
$z$ and is spatially periodic along $x$ and $y$. The initial interface between
pure light fluid and pure heavy fluid is located at $z=0$. Thus, along the $z$
axis at time $t$, the flow goes from a turbulent mixing of two fluids when $\mid
z\mid <L(t)/2$ to an intermittent border at $\mid z\mid\approx L(t)/2$ and
finally to a laminar pure fluid (light or heavy depending upon the direction)
when $\mid z\mid >L(t)/2$. The vertical velocity $u\prm_z$ of the flow is plotted
in false colour on the vertical sections. It puts to the fore large structures
going up (light gray/red) and down (dark/blue), made up of mixed fluids. The 3D bubble-like
shapes on top (at $z\approx H/2$) correspond to isosurfaces of  $u\prm_z$ at
$80\%$ of its maximum positive value.}
\label{fig:fig1}
\end{figure}

\begin{figure}[t]
\centering
   \includegraphics[width=10cm]{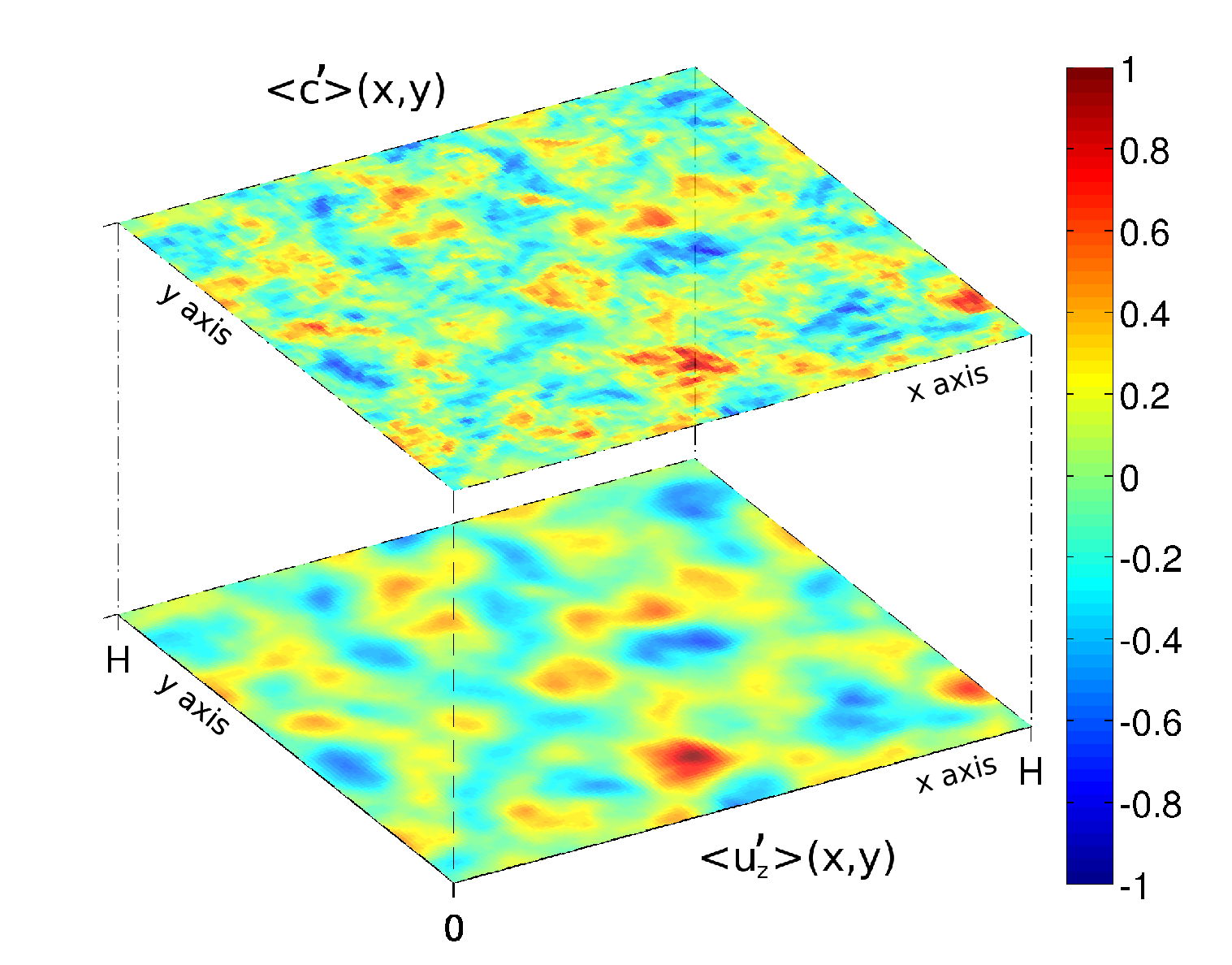}
  \caption{(Color online) Image of the {\it foliated} concentration and of the {\it foliated} vertical velocity. The top slice displays $\meanlong{c\prm}(x,y)$ in false colour at the
  instant when $L(t)=H$ for $p=0$. The bottom slice displays $-\meanlong{u_z\prm}(x,y)$
  (the minus sign is introduced so that colours match between top and bottom).
  On the colour chart, the value $1$ corresponds to the normalised maximum. The
  striking similarity in location and shape of large scales on both planes (peculiar to {\it foliated} average)
  reveals that heavy mixed fluid tends to concentrate where the flow goes down
  (dark/blue) and light mixed fluid where the flow goes up (light gray/red).} 
  \label{fig:fig2}
\end{figure}

\begin{figure}[t]
\centering
   \includegraphics[width=8cm]{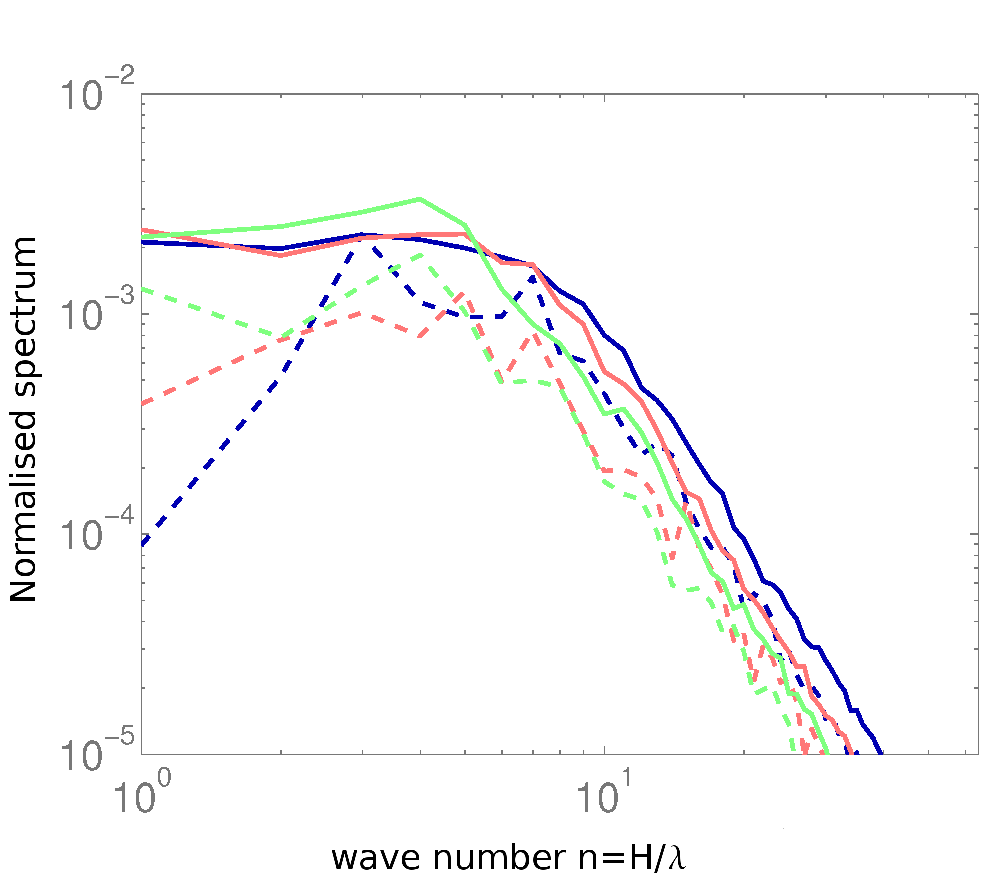}
	\includegraphics[width=8cm]{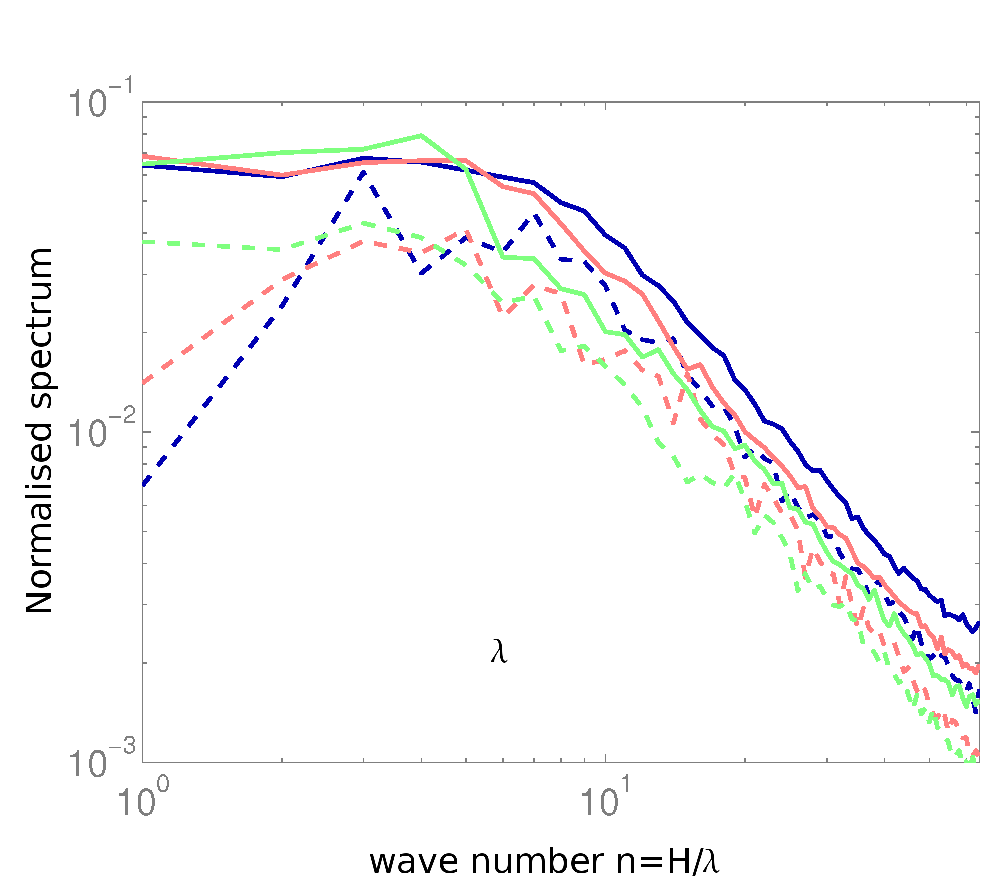}
  \caption{(Color online) Normalised {\it foliated} spectra versus transverse spectra. The left figure represents the ratio of velocity and concentration
  spectra, ${\cal E}_z(k)/\left({\cal E}_c(k)\,g(t)\,L(t)\right)$ in solid lines and $E_z(k)/
  \left(E_c(k)\,g(t)\,L(t)\right)$ in dashed lines. The right
  figure represents the ratio of production and concentration spectra, ${\cal
  E}_{cz}(k)/\left({\cal E}_c(k)\,\sqrt{g(t)\,L(t)}\right)$
  in solid lines and $E_{cz}(k)/\left(E_c(k)\,\sqrt{g(t)\,L(t)}\right)$ in dashed lines. Displayed results come from the simulation
 at constant acceleration $p=0$. The colours correspond to different times
 in the evolution of the mixing zone: $t_1$ (dark/blue)$<t_2$ (medium gray/red)$<t_3$ (light gray/green)
 defined by $L(t_1)=0.5\,H$, $L(t_2)=0.75\,H$ and $L(t_3)=H$. In
 the whole domain of wave numbers, ratios of foliated spectra are distinctively
 smoother than transverse spectra. In the domain where $n<6$ (large scales), the
 ratio of {\it foliated} spectra is constant and does not vary with time whereas, in
 the same domain, the ratio of transverse spectra can vary up to one decade.}
\label{fig:fig3}
\end{figure}

\begin{figure}[t]
\centering
   \includegraphics[width=10cm]{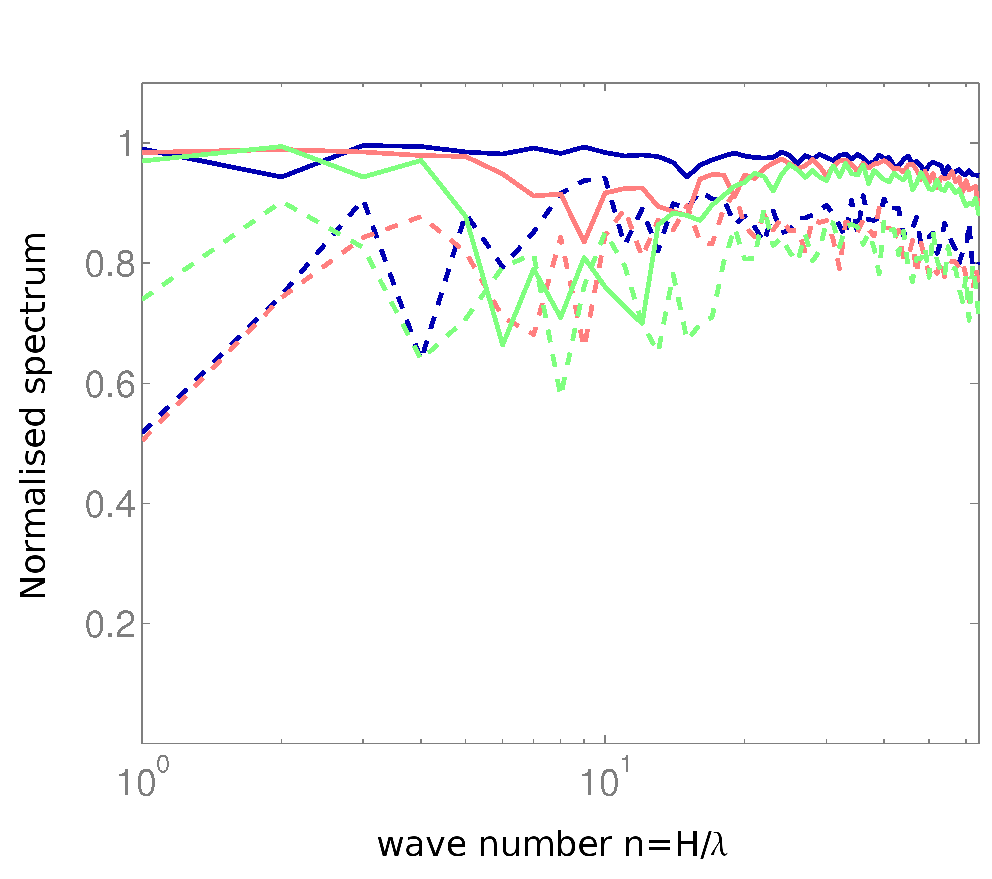}
  \caption{(Color online) Normalised correlation spectra. The evolution of ${\cal R}(k,t)=\frac{{\cal E}^2_{cz}(k,\,t)}{2\,{\cal
  E}_z(k,\,t)\,{\cal  E}_c(k,\,t)}$ (solid lines) and of
  $R(k,t)=\frac{E^2_{cz}(k,\,t)} {2\,E_z(k,\,t)\, E_c(k,\,t)}$ (dashed lines) at
  constant acceleration $p=0$ is compared at different times, $t_1$ (dark/blue) $<t_2$ (medium gray/red) $<t_3$
  (light gray/green). The simulation, at $p=0$ (but at $p=1$, $2$ and $3$ as well), shows that when $n\leq 6$, ${\cal R}=1$ to within $6\%$ and with
  the same characteristic smoothness as already depicted on Fig.\ref{fig:fig2} (vertical axis is linear). On the
  contrary, $R$ varies significantly as time goes
  by on the same range of wave numbers.}
\label{fig:fig4}
\end{figure}

\begin{figure}[t]
\centering
   \includegraphics[width=10cm]{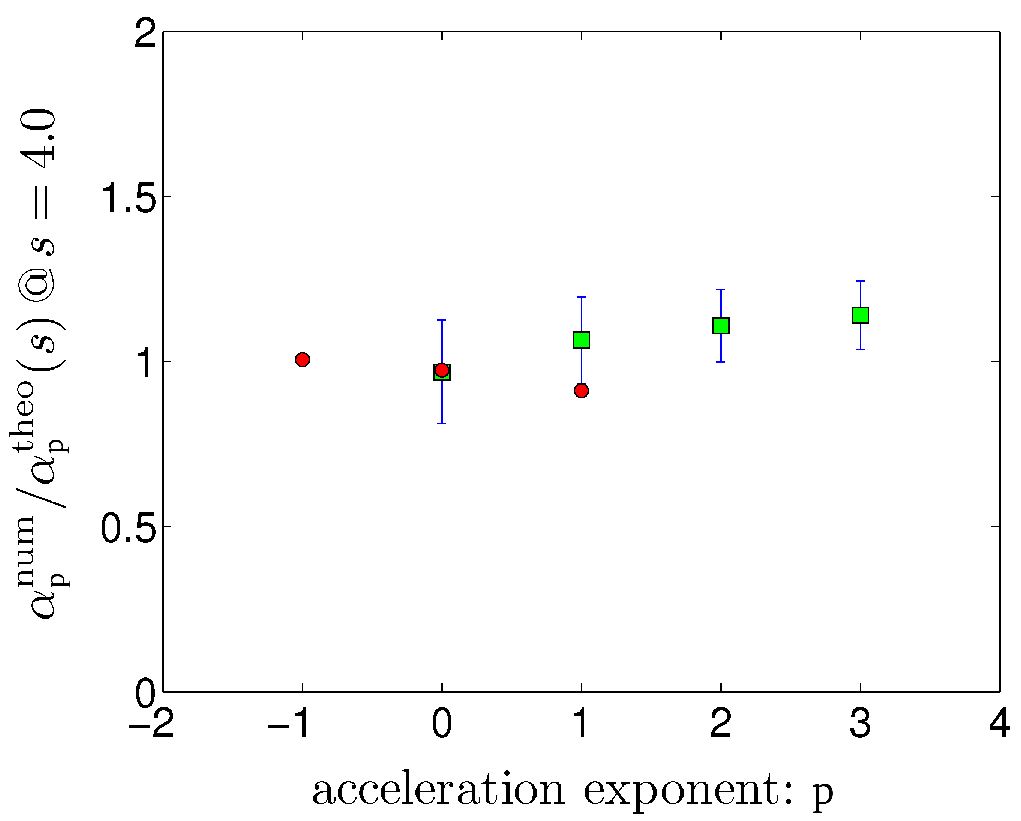}
  \caption{(Color online) Observed to predicted ratio of growth rate $\alpha_p$. Green squares correspond to results obtained
  with SURFER and performed by the authors and red circles to results performed with TURMOIL3D by D.
  Youngs and A. Llor. A least square best fit has been applied on the numerical results using the
  formula (\ref{alphan}) and it was found, assuming $s$ does not depend on $p$, that $s=4.0\pm
  0.1$. Error bars correspond
to the size of the biggest interval containing all three different measures of $\alpha_p$: (1) $\alpha_p=L(t)/({\cal A}\,g(t)\,t^2)$,
(2)  $\alpha_p=\dot{L}(t)/((p+2)\,{\cal A}\,g(t)\,t)$ and (3) the same method described in
\cite{cookcab}. Error bars were not provided in \cite{youngsllor}.}
\label{fig:fig5}
  \end{figure}

The simulations were carried out in nondimensional units. Initially, two vertically stacked fluid
layers of 
different density, such that ${\cal A}=0.1$,
have been considered (see Fig.\ref{fig:fig1}). Periodic boundary conditions
were prescribed on the four vertical domain walls, whereas no flux and
no slip conditions were imposed on the two horizontal walls of the domain,
at the top and at the bottom.
Initially, the velocity field $\bm{u}\left(x,y,z,t=0\right)$ is perturbed around the interface
using a sum of random small amplitude modes which comply with the incompressibility condition. The wave numbers selected verify $15\leq n\leq 17$ ($n\gg 1$) to get a late time self-similar evolution of the flow (mode-coupling mechanism). Kinematic viscosity $\nu$ and surface tension $\sigma$ in eq.(\ref{NS}) were chosen in such a way that they affect small scales only (large $k\approx 2\pi/\Delta$ where $\Delta$ is the mesh size). This is why ${\cal A}\,g/\Delta\approx \nu^2/\Delta^4$ and ${\cal A}\,g/\Delta\approx \sigma/(\rho\Delta^3)$. Therefore $\nu\approx \sqrt{{\cal A}\,g\,\Delta^3}$, as in \cite{alphag}, and $\sigma\approx \rho\,{\cal A}\,g\,\Delta^4$. It is important to stress that the interface reconstruction mimics the effect of a small molecular diffusion coefficient $\kappa\ll \nu$.

\par

The value of $\alpha_p$ is affected by how the width of the mixing zone, $L(t)$, is
measured or computed. Experimentally and numerically, two methods are employed and they both use
$\mean{c}(z,t)$. The threshold method prescribes $L_{\mathrm{th}}(t)=\mid z_{99}-z_1\mid$ where
$\mean{c}(z_1,t)=0.01$ and $\mean{c}(z_{99},t)=0.99$. This method is affected by statistical
fluctuations of $\mean{c}$ producing noisy $L_{\mathrm{th}}(t)$ and an even noisier derivative. The
integral method prescribes $L_{\mathrm{int}}(t)=6 \int\,\mathrm{d}z\,\mean{c}(z,t)(1-\mean{c}(z,t))$
which provides an exact value if the profile of $\mean{c}$ is linear. Since $\mean{c}$ is almost
affine everywhere in our simulation at low Atwood, the difference between $L_{\mathrm{int}}(t)$ and
$1/\partial_z\mean{c}$ is negligibly small. In a sense, the reasoning that led to the growth rate
formula dictated the definition of $L$, by the integral method, which has been used to calculate
the value of $\alpha_p$ in each of the four simulations. 

\begin{acknowledgments}
The authors would like to thank Antoine Llor for many
fruitful discussions and two anonymous referees for valuable comments.\\
$^{*}$: corresponding author's email {\tt olivier.poujade@cea.fr}.
\end{acknowledgments}

\printfigures

\end{document}